\documentstyle[12pt]{article}
\begin{document}

\baselineskip=15pt

\newcommand{\bc}{\begin{center}}
\newcommand{\ec}{\end{center}}
\newcommand{\be}{\begin{equation}}
\newcommand{\ee}{\end{equation}}
\newcommand{\bq}{\begin{eqnarray}}
\newcommand{\eq}{\end{eqnarray}}
\newcommand{\ikl}{\int_k^\Lambda}
\newcommand{\dd}{\frac{d^2k}{(2 \pi)^2}}
\newcommand{\dt}{\frac{d^3k}{(2 \pi)^3}}
\newcommand{\dtp}{\frac{d^3p}{(2 \pi)^3}}
\newcommand{\dq}{\frac{d^4k}{(2 \pi)^4}}
\newcommand{\dn}{\frac{d^nk}{(2 \pi)^n}}
\newcommand{\PLB}{{\it{Phys. Lett. {\bf{B}}}}}
\newcommand{\NPB}{{\it{Nucl. Phys. {\bf{B}}}}}
\newcommand{\PRD}{{\it{Phys. Rev. {\bf{D}}}}}
\newcommand{\AOP}{{\it{Ann. Phys. }}}
\newcommand{\MPL}{{\it{Mod. Phys. Lett. }}}
\newcommand{\del}{\partial}

\begin{titlepage}

\vskip1in

\begin{center}
\LARGE{\bf Chiral anomaly and CPT invariance in an implicit momentum space regularization framework}
\end{center}

\vskip1.0cm

{\large{A. P. B. Scarpelli$^{(1)}$, \, M.
Sampaio $^{(2)}$, \, M. C. Nemes $^{(1)}$, \, B. Hiller$^{(2)}$}}\\

\vskip1.0cm
\begin{center}
$(1)$ Federal University of Minas Gerais\\
Physics Department - ICEx\\
P.O. BOX 702, 30.161-970, Belo Horizonte MG - Brazil\\
\vskip1.0cm
$(2)$ University of Coimbra\\
Centre for Theoretical Physics\\
3004-516 Coimbra - Portugal\\
\end{center}
\vskip0.5cm
\begin{center}
{\it {carolina@fisica.ufmg.br,
msampaio@fisica.ufmg.br, scarp@fisica.ufmg.br, brigitte@malaposta.fis.uc.pt}}
\end{center}
\begin{abstract}
\noindent
This is the second in a series of two contributions in which we set out to establish a novel momentum space  framework to treat  field theoretical infinities in perturbative calculations when parity-violating objects occur. Since no analytic continuation on the space-time dimension is effected, this framework can be particularly useful to treat dimension-specific theories. Moreover arbitrary local terms stemming from the underlying infinities of the model can be properly parametrized. We (re)analyse the undeterminacy of the radiatively generated CPT violating Chern-Simons term within an extended version of $QED_4$ and calculate the Adler-Bardeen-Bell-Jackiw triangle anomaly to show that our framework is consistent and general to handle the subtleties involved when a radiative corretion is finite.
\end{abstract}
\end{titlepage}


\section{Introduction}

To circumvent the ultraviolet infinities which appear in perturbative calculations in   renormalizable quantum field theories, a relatively new regularization framework called Differential Regularization (DFR) was proposed in \cite{FREEDMAN}. DFR is a very elegant formalism and yet somewhat tractable from the calculational standpoint. The essence of this method is to write an amplitude in the position (Euclidian) space as derivatives  of a less divergent function which contains a logarithmic mass scale (playing the role of a subtraction scale) and an integration by parts prescription where a surface term is neglected. DFR has some rather nice features: a) it does not modify the dimensionality of the space-time or introduce a regulator; b) it naturally addresses the question of renormalization by delivering renormalized amplitudes which satisfy Callan-Symanzik equations. Therefore, in principle, it is applicable to a broad range of field theoretical models.

On the other hand perturbative calculations in chiral, topological and supersymmetric theories (which share the feature of being well defined only in their integer space-time dimension) are more involved since in most cases one cannot apply dimensional regularization (DR) (and some of its variants) in an ambiguity free way \cite{CS},\cite{SUSY},\cite{SM} or without having to face cumbersome calculational complications stemming from spurious anomalies. Although DFR can in principle be consistently applied to this class of theories as it has been verified in some models \cite{APPL-DFR}-\cite{PV4}, its use has not become so popular yet, especially beyond the one loop order, in the electroweak sector of the Standard Model and in SUSY models, for instance. Some authors argue that the convenience of  (momentum space) DR's beyond the one loop order justifies to mend the shortcomings caused by unwanted breakdown of symmetries \cite{SUSY},\cite{SM}. The reason is that no procedure of DFR beyond one loop order exists such that gauge invariance is automatic . Besides momentum space is more natural for calculations of amplitudes with fixed external momenta and that Feynman rules are simpler to handle, especially for massive theories (although this is of course a matter of taste), and finally that we have an all ready library of momentum space integrals, Feynman parameters, etc \footnote{DFR may also be translated into momentum space by Fourier transforming the resulting renormalized amplitude \cite{PV1} .}.

In view of this, a consistent, symmetry preserving regularization/renormalization procedure that works {\underline{directly}} in the momentum space without recoursing to the analytical continuation on the space-time dimension would be desirable and certainly worthwhile being studied.
Recently a momentum space $n$-dimensional regularization framework which shares some  advantageous features of DFR was proposed \cite{OAC},\cite{PHD},\cite{PRD},\cite{AOMC},\cite{GC},\cite{PRDOC} (for definiteness let us call it Implicit Regularization (IR)). Nonetheless, it is not a simple momentum space version of DFR: it can give us new insights in some calculations (as we shall see in this work), especially when the standard regularizations cannot be implemented, as well as understand the origin of certain regularization dependent results. In spirit, it is close to the BPHZ method: a regulating function $G(k^2,\Lambda_i)$ is only implicitly assumed in order to justify the algebraic steps in the integrands of the divergent amplitudes. $\Lambda_i$ are the parameters of the distribution $G$ for which we only assume that it is even in the integrating momentum $k$ and that a connection limit exists, i.e. $\lim_{\Lambda_i \rightarrow \infty} G(k^2,\Lambda_i)=1$, so to guarantee that the finite amplitudes are not modified\footnote{Throughout the paper we write $\int^\Lambda_k F(k,p)$ for $\int_k F(k,p) G(k^2,\Lambda_i)$ .}. The purpose is to display the divergences in terms of primitive divergent integrals which depend solely on the loop momenta which need not be evaluated.  The remaining finite integrals can be grouped in two classes: the ones which depend  on the physical momenta, and therefore are integrated as usual, and differences between integrals of the same degree of divergence $\Delta_s$. The latter, which we have  called consistency relations (CR) in early works, will play a central role in our discussion in this contribution. For any (integer) space-time dimension such CR will appear systematically in perturbative calculations in any field theoretical model. Their value is regularization dependent and therefore, in a most general formulation, undetermined. However it may be fixed by the symmetries of the underlying theory (Ward identities) at the very final stage of the calculation. 

A crucial feature of the regularization framework described above is that the singular terms are left untouched in the form of basic divergent integrals. No finite terms are lost whereas arbitrary local terms may be parametrized by the CR. According to renormalization theory such arbitrary local terms correspond to the addition of a finite counterterm in the Lagrangian, which may be added at will as long as it respects the relevant symmetries of the theory. Again, IR and DFR partake these feature and therefore give rise to the most general quantum effective action since the CR play the role of the arbitrary mass scale that appears in DFR. This is precisely what distinguishes DFR and IR from DR, Pauli-Villars etc., which together with a renormalization prescription,  fix these arbitrary local terms ab initio \cite{CHENCPT}.

A constrained version of IR (CIR) where the CR are set to zero would  be more  practical from the calculational point of view.  It  amounts to fixing some arbitrary scales from the start in such a way that the Ward Identities are preserved \cite{OAC}.  Whereas this could be advantageous for, say, all order proofs, care must be exercised when
one computes amplitudes with (an odd number of) parity violating objects, such as $\gamma_5$ matrices. That is because it  can be shown that such CR are connected to momentum routing invariance in the loop of a Feynman diagram. Should the CR vanish then the amplitude is momentum routing invariant \footnote{However if they assume a non-vanishing value, it does not necessarily mean that momentum routing invariance is broken.} \cite{OAC},\cite{PRD}. In particular, they are important to study chiral theories and chiral anomalies since momentum routing dependence plays a key role in describing chiral anomalies within perturbation theory \cite{JACKIWCA} which  is the main subject of this contribution. The counterpart in DFR is called Constrained Differential Regularization (CDFR) program \cite{PV1}-\cite{PV4}, in which all the arbitrary scales are fixed (except for the renormalization scale) by means of a set of rules.

This paper is organized as follows: In section \ref{s:int} we recall the main features of  IR and present the CR, using $QED_4$ as an illustration. The novelty here is that we introduce a general parametrization for the arbitrary terms which is consistent with the symmetric limit in the integration variable of divergent integrals. In section \ref{s:cpt} we briefly revisit the problem of the radiatively induced $CPT$ and Lorentz violation within an extended version of $QED_4$, namely adding a term of the form $\bar \psi \gamma_5  b\hspace{-2.0mm}/\psi$ in the Lagrangian ($b_\mu$ is a constant four-vector). This has been a matter of intensive debate as regards of the role played by the regularization within a perturbative or a nonperturbative in $b$ treatment of the problem. Here we show explicitly that the undeterminacy arises in both treatments in the same fashion in IR. In section \ref{s:abj}, it is shown how IR can be used to consistently display the traingle chiral anomaly in a scheme-free fashion which allows the anomaly to appear in the vector and axial Ward identites on equal footing.  In all the examples we emphasize the role played by the choice of local arbitrary terms in our framework, whose value is either determined by the symmetries of the underlying theory or, if not, should be left arbitrary.

\section{Arbitrary local terms and momentum routing}
\label{s:int}
Let $k$ be the momentum running in a loop of a Feynman diagram. In \cite{OAC}, \cite{PRD} it was shown that if certain well defined differences between divergent integrals \footnote{From now on we simply refer to them as Consistency Relations (CR)}(which do not depend on external momenta and have identical $\Delta_s$) were to vanish then the corresponding  amplitude is independent of the arbitrary momentum routing in a loop, consistently with energy-momentum conservation. It is easy to see that {\it all} the CR below vanish identically if we perform an explicit evaluation within dimensional or Pauli-Villars regularization \footnote{The linear and quadratic CR's are not finite in all regularization schemes (e.g. naive cut-off in the $\Lambda \rightarrow \infty$ limit). This is not a problem because a scheme in which they are finite can be found and in non-anomalous
situations a local contraterm can be added to yield zero. }.
 
They can be grouped according to the space-time dimension \cite{PRD}:
\begin{itemize}
\item{ \bf{1+1 Dimensions} }
\be
\Delta_{\mu \nu}^0 \equiv \int^{\Lambda}
\frac{d^2k}{(2\pi)^2}\frac{g_{\mu\nu}}{k^2-m^2}-2\int^{\Lambda}
\frac{d^2k}{(2\pi)^2}\frac{k_{\mu}k_{\nu}}{(k^2-m^2)^2},
\label{CR2}
\ee
\item{ \bf{2+1 Dimensions} }
\be
\Xi_{\mu \nu}^1 \equiv \int^{\Lambda}
\frac{d^3k}{(2\pi)^3}\frac{g_{\mu\nu}}{k^2-m^2}-
2\int^{\Lambda} 
\frac{d^3k}{(2\pi)^3}\frac{k_{\mu}k_{\nu}}{(k^2-m^2)^2},
\label{CR31}
\ee
\be
\Xi_{\mu \nu \alpha \beta}^1 \equiv
g_{\{\mu\nu}g_{\alpha\beta \}}
\int^{\Lambda} \frac{d^3k}{(2\pi)^3}\frac{1}{k^2-m^2}
-8\int^{\Lambda} 
\frac{d^3k}{(2\pi)^3}\frac{k_{\mu}k_{\nu}k_{\alpha}k_{\beta}}{(k^2-m^2)^3},
\label{CR32}
\ee
etc..     
\item{ \bf{3+1 Dimensions} }
\be
\Upsilon_{\mu \nu}^2 \equiv \int^{\Lambda}
\frac{d^4k}{(2\pi)^4}\frac{g_{\mu\nu}}{k^2-m^2}-
2\int^{\Lambda} 
\frac{d^4k}{(2\pi)^4}\frac{k_{\mu}k_{\nu}}{(k^2-m^2)^2},
\label{CR4Q1}
\ee
\be
\Upsilon_{\mu \nu}^0 \equiv \int^{\Lambda}
\frac{d^4k}{(2\pi)^4}\frac{g_{\mu\nu}}{(k^2-m^2)^2}-
4\int^{\Lambda} 
\frac{d^4k}{(2\pi)^4}\frac{k_{\mu}k_{\nu}}{(k^2-m^2)^3},
\label{CR4L1}
\ee
\be
\Upsilon_{\mu \nu \alpha \beta}^2 \equiv
g_{\{\mu\nu}g_{\alpha\beta \}}
\int^{\Lambda} \frac{d^4k}{(2\pi)^4}\frac{1}{k^2-m^2}
-8\int^{\Lambda} 
\frac{d^4k}{(2\pi)^4}\frac{k_{\mu}k_{\nu}k_{\alpha}k_{\beta}}{(k^2-m^2)^3},
\label{CR4Q2}
\ee
\be
\Upsilon_{\mu \nu \alpha \beta}^0 \equiv
g_{\{\mu\nu}g_{\alpha\beta \}}
\int^{\Lambda}
\frac{d^4k}{(2\pi)^4}\frac{1}{(k^2-m^2)^2}
-24\int^{\Lambda} 
\frac{d^4k}{(2\pi)^4}\frac{k_{\mu}k_{\nu}k_{\alpha}k_{\beta}}{(k^2-m^2)^4},
\label{CR4L2}
\ee
\end{itemize}
etc., where $g_{\{\mu\nu}g_{\alpha\beta \}}$ stands for $g_{\mu\nu}g_{\alpha\beta}+g_{\mu\alpha}g_{\nu\beta}+g_{\mu\beta}g_{\nu\alpha}$. 

On the other hand it is well known that a shift in $k$ is immaterial only if $\Delta_s \le 0$, otherwise  a (finite) surface term should be added. This is an indication that care must be exercised in what concerns the momentum routing when divergences higher than logarithmic arise in Feynman diagram calculations. Perturbation theory makes a peculiar use of this feature for in some cases gauge invariance relies on adopting a special momentum routing \cite{JACKIWCA}. 

The most famous example is the triangle chiral anomaly (see for instance \cite{JACKIWCA}). It is noteworthy that the amplitudes which manifest this feature generally contain one axial vertex (parity violating object). A closely related issue is that whilst a shift in the integration variable is allowed within dimensional regularization, the algebraic properties of $\gamma_5$ clash with analytical continuation on the space-time dimension. This suggests that working with CIR in the presence of dimension-specific objects may give rise to similar problems as those appearing in dimensional reduction beyond the one loop order \cite{SIEG}, \cite{WIP}.

We shall work in a regularization framework where a regulator needs only implicitly be assumed as discussed in the introduction. For a more detailed account on IR please see \cite{OAC}. The basic procedure is to isolate the divergences of the amplitudes in the form of basic divergent integrals namely,
\be
I_{quad}^{0 \, \Lambda} (m^2) = \int^\Lambda \frac{d^4 k}{(2 \pi)^4} \frac{1}{k^2-m^2} 
\label{IQ4}
\ee
\be
I_{log}^{ 0 \, \Lambda} (m^2) = \int^\Lambda \frac{d^4 k}{(2 \pi)^4} \frac{1}{(k^2-m^2)^2} 
\label{IL4}
\ee
\be
I_{lin}^{ 0 \, \Lambda} (m^2) = \int^\Lambda \frac{d^3 k}{(2 \pi)^3} \frac{1}{k^2-m^2} 
\label{IL3}
\ee
and so on, which carry no dependence on the external momenta. The latter will appear only in finite integrals. This can be achieved by means of a convenient identity at the level of the integrand, namely
\be
\frac{1}{[(k+k_i)^2-m^2]}=\sum_{j=0}^N\frac{\left(
-1\right) ^j\left( k_i^2+2k_i\cdot k\right) ^j}{\left(
k^2-m^2\right) ^{j+1}}
+\frac{\left( -1\right) ^{N+1}\left( k_i^2+2k_i\cdot
k\right) ^{N+1}}{\left(
k^2-m^2\right) ^{N+1}[ \left( k+k_i\right) ^2-m^2] },
\label{rr}
\ee
where $k_i$ are the external momenta and $N$ is chosen so that the last term is finite under integration over $k$. The basic divergent integrals as defined in (\ref{IQ4}),(\ref{IL4}) and (\ref{IL3}) (and which can be used to characterize the divergent structure of the underlying model) need not be evaluated: they can be fully absorbed in the definition of the renormalization couplings. For the evaluation of the renormalization group $\beta$-function to two loop order in $\varphi^4_4$-theory and $QED_4$ within this approach see \cite{AOMC}. For an algebraic proof of renormalizability to $n$-loop order of $\varphi^3_6$ theory (alternative to the BPHZ method) see \cite{GC}. Applications to models involving parity-violating objects can be found in \cite{PRD} and to non-renormalizable theories in \cite{PRDOC}.

In order to get some insight into our discussion let us display the $QED_4$ vacuum polarization tensor according to the rules of IR. To one loop order and with arbitrary momentum routing it reads \footnote{Throughout this paper $\int_k$ stands for $\int \frac{d^4 k}{(2 \pi)^4}$.}
:
\be
\Pi_{\mu \nu }= - \int_k
\mbox{tr} \left\{ \gamma _\mu S(k+k_1)\gamma _\nu
S(k+k_2)\right\} ,
\label{vptqed}
\ee
where $S(k)$ is the usual free fermion propagator. For the particular momentum routing $k_1=p$ and $k_2=0$, (\ref{vptqed}) can be written, after taking the trace over the Dirac matrices, as 
$$
\Pi_{\mu \nu }= - 4 ( 2 J_{\mu \nu}^\Lambda + p_\mu J_\nu^\Lambda +  p_\nu J_\mu^\Lambda - g_{\mu \nu}I^\Lambda)
$$
where
\be
J^\Lambda, J_\mu^\Lambda, J_{\mu \nu}^\Lambda \equiv \int_k^\Lambda\frac{1, k_\mu, k_\mu k_\nu}{[(k+p)^2-m^2](k^2-m^2)}, 
\ee
\be
I^\Lambda \equiv \int_k^\Lambda \frac{k^2-m^2+p\cdot k}{[(k+p)^2-m^2](k^2-m^2)} \, .
\ee
Notice that the integrals above are divergent. According to our strategy we display the divergencies solely as a function of the loop momentum by means of a recursive use of (\ref{rr}) to yield:
\bq
&&\frac{\Pi_{\mu \nu }}{4} = 2 \ikl \frac{k_\mu k_\nu}{(k^2-m^2)^2}- g_{\mu \nu} \ikl \frac{k^2}{(k^2-m^2)^2} + m^2 g_{\mu \nu} \ikl \frac{1}{(k^2-m^2)^2} - \nonumber \\
&&- p^2
\ikl \frac{2 k_\mu k_\nu}{(k^2-m^2)^3} + 8 p^\alpha p^\beta \ikl \frac{k_\mu k_\nu k_\alpha k_\beta}{(k^2-m^2)^4} - 2 p^\alpha p_\nu \ikl \frac{k_\alpha k_\mu}{(k^2-m^2)^3}\nonumber \\
&& - 2 p^\alpha p_\mu \ikl \frac{k_\alpha k_\nu}{(k^2-m^2)^3} - p^2 g_{\mu \nu} \ikl \frac{k^2}{(k^2-m^2)^3}- 4 g_{\mu \nu} p_\alpha p_\beta \ikl \frac{k^2 k_\alpha k_\beta}{(k^2-m^2)^4} + \nonumber \\
&& 2 g_{\mu \nu} p_\alpha p_\beta \ikl \frac{k_\alpha k_\beta}{(k^2-m^2)^3} -  \frac{b}{3} \Big( p^2g_{\mu
\nu }-p_\mu p_\nu \Big)\Bigg( \frac{1}{3}+\frac{(2m^2+ p^2)}{p^2}
Z_0(p^2;m^2)\Bigg) 
\label{qedinf}
\eq
where $Z_0 (p^2;m^2)$ is defined as in (\ref{zk}) and 
$$
b \equiv \frac{i}{(4 \pi)^2}.
$$
The last term in (\ref{qedinf}) is the result of the integration of finite integrals. To define the renormalized vacuum polarization tensor one should join the usual counterterm  to define ${\Pi}_{\mu \nu R}= {\Pi}_{\mu \nu} + (p_\mu p_\nu - p^2 g_{\mu \nu})(Z_3 - 1)$, $A^\mu = Z_3^{1/2} A^\mu_R$.  As it is well known the Ward identities strongly constrain the divergent structure, namely the infinity ought to be absorbed by $Z_3$. 

In the spirit of our method, namely to write the infinite parts in terms of the loop momenta $k$ only, we could proceed in two equivalent ways: 1) To write a parametrization based on very general properties of the divergent integrals and 2) To group differences of integrals of same $\Delta_s$ to define the objects expressed in eqns. (\ref{CR2})-(\ref{CR4L2}). The purpose which is common to both approaches is to make close contact with Jackiw's idea that the arbitrariness expressed by finite differences between divergent integrals should be left for the symmetries of the underlying model to fix \cite{JACKIWFU}. Although we are evidently more interested in the second approach as we have discussed in the introduction, it will be also interesting to analyse the problem of the CPT violation in extended $QED_4$, (section \ref{s:cpt}), in the light of the first approach.

For this purpose, we write a general parametrization for divergent integrals which depend only on $k$. For instance, consider (\ref{IL4}). It can be shown that
\be
\frac{\del I_{log}^{0 \, \Lambda}(m^2)}{\del m^2}= - \frac{b}{m^2}. 
\ee
from which we see that 
\be
\tilde I_{log}^{\Lambda}(m^2) = b \ln \Big( \frac{\Lambda^2}{m^2}  \Big) + \beta \, ,
\label{pil}
\ee
where $\beta$ is a finite constant and $\Lambda$ is a cut-off, is a general parametrization of (\ref{IL4}). In fact, also for a generic logarithmic divergence $I_{log}^{i \, \Lambda} (m^2)$ 
\be
\frac{ \del I_{log}^{i \, \Lambda} (m^2)}{\del m^2} = \frac{\del }{\del m^2}\ikl \frac{k^{2i}}{(k^2-m^2)^{i+2}} = - \frac{b}{m^2} \, ,
\ee
and thus (\ref{pil}) is its general parametrization. For the quadratic divergences, we write
\be
I_{quad}^{i \, \Lambda} (m^2) = \ikl \frac{k^{2i}}{(k^2-m^2)^{i+1}} \,
\ee
in which $i=0$ corresponds to our basic quadratic divergent (\ref{IQ4}). However, in this case different values of $i$ will render different parametrizations for the $I_{quad}^{i \, \Lambda} (m^2)$ as
\be
\frac{ \del I_{quad}^{i \, \Lambda} (m^2)}{\del m^2} = (i+1) I_{log}^{i \, \Lambda} (m^2)\, .
\ee
Using (\ref{pil}) and integrating the equation above, we conclude that
\be
\tilde I_{quad}^{i \, \Lambda} (m^2) =  b (i+1) \Big(c \Lambda^2 +  m^2 \ln \Big( \frac{\Lambda^2}{m^2} \Big) + \alpha m^2 \Big)\, ,
\label{piq}
\ee
where $\alpha$ and $c$  are undetermined constants, parametrizes $I_{quad}^{i \, \Lambda}(m^2)$. \footnote{After a redefinition of variables which includes $c \Lambda^2 = \tilde \Lambda^2$, we can rewrite (\ref{piq}) as 
$$
\tilde I_{quad}^{i \, \tilde \Lambda} (m^2) =  b (i+1) \Big(\tilde \Lambda^2 +  m^2 \ln \Big( \frac{\tilde \Lambda^2}{m^2} \Big) + \alpha' m^2 \Big),
$$
keeping in mind that $\Lambda$'s coming from different divergent pieces should be labelled.}
Notice that such parametrizations are based on most general properties of the primitive divergent integrals. One can find general parametrizations for other divergencies in any space-time dimension in a similar fashion. For the sake of illustration let us test our parametrizations against two CR. For example (\ref{CR4L1}) reads
\be
\Upsilon_{\mu \nu}^0 = g_{\mu \nu}\Big(  I_{log}^{0 \, \Lambda} (m^2) - 4\, c \, I_{log}^{1 \, \Lambda} (m^2)\Big),
\label{ups0}
\ee
where we used $\int_k f(k^2) k_\mu k_\nu = c \, \int_k f(k) k^2 g_{\mu \nu}$. According to (\ref{pil}), we can parametrize (\ref{ups0}) as
\be
\tilde \Upsilon_{\mu \nu}^0 = g_{\mu \nu} b \Bigg( \ln \Big( \frac{\Lambda_1^2}{m^2} \Big) - 4 c \, \ln \Big( \frac{\Lambda_2^2}{m^2}\Big) + \beta_1 - \beta_2 \Bigg),
\ee
from which we see that the symmetric limit $c=1/4$ may be taken to yield a finite undetermined constant expressed by the difference $\beta_1 - \beta_2$.  Likewise (\ref{CR4Q1}) reads
\be
\Upsilon^2_{\mu \nu} = g_{\mu \nu} \Big(  I_{quad}^{0 \, \Lambda} (m^2) - 2\, c \, I_{quad}^{1 \, \Lambda} (m^2)\Big),
\ee
for which  (\ref{piq}) allows us to write
\be
\tilde \Upsilon^2_{\mu \nu} = b g_{\mu \nu} \Bigg(\Lambda_1^2 + m^2 \ln \Big( \frac{\Lambda_1^2}{m^2} \Big) + \alpha_0 m^2 - 4 \, c \Big( \Lambda_2^2 +  m^2 \ln \Big( \frac{\Lambda_2^2}{m^2} \Big) + \alpha_1 m^2 \Big) \Bigg)\, 
\ee
which explicitly shows that the symmetric limit $c = 1/4$ is necessary to ensure the finiteness of $\tilde \Upsilon^2_{\mu \nu}$ provided $\Lambda_1^2 =  \Lambda_2^2$.

In other words, at least within our parametrization, the symmetric limit can be consistently taken and the Ward identities are ultimately used to fix the ambiguities.  In the light of what we have just exposed let us turn our attention back to the $QED_4$ vacuum polarization tensor. Except for the first three integrals in (\ref{qedinf}), the other logarithmically divergent integrals can be added up to yield a gauge invariant structure of divergence:
\bq
\Pi_{\mu \nu }^R &=&4 \Bigg( 2 \ikl \frac{k_\mu k_\nu}{(k^2-m^2)^2}- g_{\mu \nu} \ikl \frac{k^2}{(k^2-m^2)^2} + m^2 g_{\mu \nu} \ikl \frac{1}{(k^2-m^2)^2} \Bigg) \nonumber \\
&+& \frac{4}{3} ( p^2 g_{\mu \nu }-p_\mu p_\nu )\Bigg( \tilde I_{log}^{\Lambda}(m^2)- b \Big(\frac{1}{3}+\frac{(2m^2+ p^2)}{p^2} Z_0(p^2;m^2)\Big) \Bigg) \nonumber\\
&+&  (p^2g_{\mu \nu }-p_\mu p_\nu) (Z_3 - 1).
\label{qedvpt2}
\eq
The Ward identities impose that the first three integrals in (\ref{qedvpt2}) should cancel out. We can explicitly verify within our parametrization that this amounts to saying that a finite arbitrary term should vanish. Substituting $k_\mu k_\nu$ with  $c \, k^2 g_{\mu \nu}$ in the integrand of the first integral we have, according to our notation,
\be
2\, c \, g_{\mu \nu} I_{quad}^{1 \, \Lambda_1} (m^2) - g_{\mu \nu }I_{quad}^{1 \, \Lambda_2} (m^2) + m^2 I_{log}^{0 \, \Lambda_3} (m^2) 
\ee
which can be parametrized as
\be
4 \, c \, b \, g_{\mu \nu} \Big( \Lambda_1^2 + m^2 \ln \Big( \frac{\Lambda_1^2}{m^2}\Big) + \alpha_1 m^2\Big) - 2 \, b \, g_{\mu \nu} \Big( \Lambda_2^2 + m^2 \ln \Big( \frac{\Lambda_2^2}{m^2}\Big) + \alpha_2 m^2\Big) + m^2 \, g_{\mu \nu}\, b  \Big( \ln \Big( \frac{\Lambda_3^2}{m^2}\Big) + \alpha_3 \Big) 
\ee
and hence the quadratic divergences cancel out for $2 c \Lambda_1^2 = \Lambda_2^2 $ as well as the logarithmic ones provided $c = 1/4$. Then a remaining finite term is set to zero on gauge invariance grounds.

Alternatively, and more elegantly, we can display the divergent structure of (\ref{vptqed}) in terms of the CR as expressed by (\ref{CR4Q1})-(\ref{CR4L2}) and thus learn how gauge symmetry interplays with momentum routing. We quote the result from \cite{PRD}
\bq
\Pi_{\mu \nu } &=& \tilde\Pi_{\mu \nu } +
4\Bigg(\Upsilon^2_{\mu\nu}-\frac{1}{2}(k_1^2+k_2^2)\Upsilon^0_{\mu
\nu}
+\frac{1}{3}(k_{1}^{\alpha}k_{1}^{\beta}+k_{2}^{\alpha}k_{2}^{\beta}
+k_{1}^{\alpha}k_{2}^{\beta})
\Upsilon^0_{\mu \nu \alpha 
\beta} \nonumber \\ &-&
(k_1+k_2)^{\alpha}(k_1+k_2)_{\mu}\Upsilon^0_{\nu
\alpha} 
-\frac{1}{2}(k_1^{\alpha}k_1^{\beta}+k_2^{\alpha}k_2^{\beta})g_{\mu \nu}
\Upsilon^0_{\alpha \beta}  \Bigg)\,\,\,\, \mbox{where}
\label{qedvp}
\eq
\vspace{-0.5cm}
\bq
\tilde\Pi_{\mu \nu } &=& \frac{4}{3} \Big(
(k_1-k_2)^2g_{\mu
\nu }-(k_1-k_2)_\mu (k_1-k_2)_\nu \Big) \times 
\nonumber \\
&\times& \Bigg( I_{log}^\Lambda (m^2) - b \Bigg( \frac
13+\frac{(2m^2+\left( k_1-k_2\right) ^2)}{\left(
k_1-k_2\right) ^2}
Z_0(\left( k_1-k_2\right) ^2;m^2)\Bigg) \Bigg) \, .
\eq
We can certainly set all  $\Upsilon$'s to zero consistently with the Ward identity $(k_1 - k_2)^\mu \Pi_{\mu \nu} = 0$. In addition one may choose a particular routing, say $k_1= p$, $k_2 = 0$, and let the value assumed by the $\Upsilon$'s  be arbitrary, viz. $\Upsilon^0_{\mu \nu \alpha \beta} = \lambda_1 g_{\{ \mu \nu}g_{\alpha \beta \}}$, $\Upsilon^0_{\mu \nu} = \lambda_2 g_{\mu \nu}$, $\Upsilon^2_{\mu \nu} = \lambda_3 \mu^2 g_{\mu \nu}$. Hence we obtain that
$$
p^\mu \Pi_{\mu \nu} = 4 p_\nu ( (\lambda_1 - 2 \lambda_2)p^2 + \lambda_3 \mu^2)
$$ 
from which we see that gauge invariance is accomplished through the choice $(\lambda_1,\lambda_2,\lambda_3)=(2 \lambda,\lambda,0)$, with $\lambda$ being an arbitrary local term.
The momentum routing is immaterial in this example. 

Besides being elegant, to write the differences between divergent integrals with no external momentum dependence and with the same $\Delta_s$ in terms of the CR's will be particularly attractive in the discussion of chiral theories and anomalies within perturbation theory.

\section{Undeterminacy of the induced Lorentz and CPT symmetry breaking term in extended $QED_4$}
\label{s:cpt}

The subject of a possible radiative generation of the Lorentz and CPT symmetry breaking term 
\be
{\cal{L}}_{CS} = \frac{1}{2} c_\mu \epsilon^{\mu \nu
\lambda \rho} F_{\nu \lambda}A_\rho \, , 
\label{lcptbt}
\ee
arising from  the Lorentz and CPT violating term $b_\mu
\bar{\psi}\gamma_\mu \gamma_5 \psi$ added in the fermionic
sector of standard $QED_4$ ($c_\mu$, $b_\mu$ are constant $4$-vectors) has been a matter of fiery debate. The controversy results from two main points: 1) the fact the that value of $c_\mu$ appears to be regularization dependent , \cite{CHENCPT} \cite{CPTPV}, \cite{CPTDIM} and 2) whether or not gauge invariance (and analiticity) can constrain $c_\mu$ to vanish \cite{COL}, \cite{BON},\cite{JACPT},\cite{PER2}. As firstly pointed out by Jackiw and Kosteleck\'y \cite{JACPT} and well-argued by P\'{e}rez-Victoria \cite{PER2}, the answer to this problem lies upon two related facts. Firstly, it is the action correspondent to the induced density (\ref{lcptbt}), $\int d^4 x \, {\cal{L}}_{CS}$, which is gauge invariant. Secondly, $b_\mu$ is a constant field and therefore the term which is sought after needs to be gauge invariant at zero external momentum only. Usually the  regularizations which have been employed enforce gauge invariance at all axial momenta (Pauli-Villars, DR with the 't Hooft Veltman prescription for $\gamma_5$). Other generalizations of $\gamma_5$ in DR often give different results. DFR can be employed in the perturbative in $b_\mu$ calculation \cite{CHENCPT}. However for the $b_\mu$-exact propagator case DFR cannot be implemented since it is not possible to write such propagator in coordinate space. This issue is particularly attractive to be dealt with within IR. 

Although in principle (\ref{lcptbt}) is a local renormalizable term allowed by the symmetries of the theory (save Lorentz and CPT), its appearance at classical level
undergoes stringent theoretical and experimental bounds \cite{BOUNDS}. The quantity of interest for deciding whether  (\ref{lcptbt}) is radiatively generated is the $O(A^2)$ part of 
\be
\Gamma (A^2) = -i {\mbox{ln  det}}(i \partial
\hspace{-2.0mm}/ - A\hspace{-2.0mm}/ -
b\hspace{-2.0mm}/ \gamma^5 -m )\Big|_{A^2} \propto \int_k A^\mu(-k)\Pi_{\mu \nu}(k)A_{\nu}(k).
\ee
with $\Pi_{\mu \nu} \sim b_\alpha \Gamma^{\mu \nu
\alpha}(p, -p)$. On general grounds we can say that $\Gamma^{\mu \nu \alpha}(p, -p)$ \footnote{ Diagrammatically it corresponds to a triangle graph composed of two vector currents and one axial vector current with zero momentum transfer between the two vector gauge field vertices.} is
undetermined by an arbitrary parameter $a$, namely $
\Gamma^{\mu \nu \alpha}(p, -p) \sim \Gamma^{\mu \nu
\alpha}(p, -p) + 2 i a \epsilon^{\mu \nu \alpha \beta}
p_\beta \, $ which, contrarily to the triangle anomaly, cannot be fixed by the Ward identities.

In this section we would like to recall briefly some of our results on this matter \cite{PRD} and shed some light on the interpretation, particularly in what concerns the undeterminacy of the radiatively generated term. In our framework it will become apparent that both the perturbative and non-perturbative in $b_\mu$ formulation (as conceived in \cite{JACPT}) deliver an equally intrinsically undetermined result (in the sense that different regularizations would assign different values). Thus it should be fixed by (re)normalization conditions and/or direct comparison with experiment. 

A non-perturbative evaluation makes use of the $b_\mu$-exact propagator
\be
S'(k)=
\frac{i}{ik\hspace{-2.3mm}/-m-b\hspace{-2.0mm}/\gamma_{5}}
\label{exprop}
\ee
and was thought to lead to a well determined result \cite{JACPT},\cite{CHUNG1}. The calculation is based upon the fact that (\ref{exprop}) can be decomposed as 
\be
S'(k)=S_F(k)+S_b(k),
\ee
where $S_F(k)$ is the usual free fermion propagator
and
\be
S_b(k)=
\frac{1}{ik\hspace{-2.0mm}/-m-b\hspace{-2.0mm}/\gamma_{5}}
b\hspace{-2.0mm}/ 
\gamma_{5}S_F(k).
\ee
whereas the vacuum polarization tensor can be
generically written as in \cite{JACPT}
\be
\Pi^{\mu \nu}=\Pi^{\mu \nu}_0 + \Pi^{\mu \nu}_b +
\Pi^{\mu \nu}_{bb}.
\label{tva}
\ee
The $b_\mu$-linear contribution to (\ref{lcptbt}) comes from $\Pi^{\mu \nu}_b$,
\be
\Pi^{\mu \nu}_b(p)=\int_k\mbox{tr}
\left\{\gamma^{\mu}S_F(k)
\gamma^{\nu}S_b(k+p) + \gamma^{\mu}S_b(k)
\gamma^{\nu}S_F(k+p) \right\}.
\ee 
It is remarkable that only the lowest order in $b_\mu$ approximation of $S_b(k)$, viz. $S_b(k) \sim
-iS_F(k)b\hspace{-2.0mm}/\gamma_5 S_F(k)$ (and thus $\Pi^{\mu \nu}_b=\Pi^{\mu \nu \alpha}b_{\alpha}$), coincides with the result to all
orders \footnote{This fact combined with the regularization ambiguity of $c_\mu$ has motivated an analogy with the triangle anomaly in a model calculation in which $b_\mu$ can  be initially considered as a non-constant field $b_\mu(x)$ and $CPT$ is spontaneously broken \cite{PER2}.}\cite{CHUNG1},\cite{CHUNG2},\cite{PER1},\cite{PER2}. Therefore we need to evaluate
\bq
\Pi^{\mu \nu \alpha}(p)&=& -i
\int_k \mbox{tr} 
\left\{\gamma^{\mu}S(k)
\gamma^{\nu}S(k+p)\gamma^{\alpha}\gamma_5 S(k+p)
 + \gamma^{\mu}S(k)\gamma^{\alpha}\gamma_5 S(k )
\gamma^{\nu}S(k+p) 
\right\} \nonumber
 \\
&\equiv&-\left\{I_1^{\mu \nu \alpha} + I_2^{\mu \nu
\alpha} \right\}
\eq
which, within IR, yields (see \cite{PRD} for calculational details)
\be
I_2^{\mu\nu\alpha}=I_1^{\mu\nu\alpha}+\frac{\lambda}{2
\pi^2}p_{\beta}
\epsilon^{\mu\nu\alpha\beta},
\ee  
where $\lambda$ is an undetermined dimensionless parameter defined as in (\ref{CR4L1}) with 
\be
\Upsilon^0_{\alpha \beta} = \frac{i \lambda}{8 \pi^2} g_{\alpha \beta}
\label{undcpt}
\ee
and $I_1^{\mu\nu\alpha}$ is finite, and can be readily evaluated to give
\be
\Pi^{\mu\nu\alpha}_{non-pert}= 
\epsilon^{\mu\nu\alpha\beta}\frac{p_{\beta}}{2 \pi^2}
\Bigg( \frac{\theta}{\sin \theta}- \lambda \Bigg) \, ,
\label{cptnp}
\ee  
where $\theta = 2 {\mbox{arcsin}}(\sqrt{p^2}/(2 m))$
and $p^2 < 4 m^2$. As for the perturbative in $b_\mu$ calculation, the relevant diagrams are the
$b_\mu$-linear one loop correction to the photon
propagator in which a factor $i b_\lambda
\gamma^\lambda \gamma^5$ can be inserted in either of
the two internal fermionic lines to render equal
contributions. Thus the amplitude reads
\be
\Pi_{\mu \nu}^b = 2 \, (-i) b^\lambda \int_k
{\mbox{tr}} \gamma_\mu S_F (k-p) \gamma_\nu S_F (k)
\gamma_\lambda \gamma^5 S_F (k) \equiv 2 b^\lambda
\Pi_{\lambda \mu \nu}\, ,
\ee
where $p$ is the external momentum. The integral above
is just our $I^{\mu \nu \alpha}_2$ in the
non-perturbative with $p \rightarrow -p$ and the
$\mu$, $\nu$ indices interchanged. Taking into account the change of signs, it gives
\be
\Pi^{\mu\nu\alpha}_{pert}= 
\epsilon^{\mu\nu\alpha\beta}\frac{p_{\beta}}{2 \pi^2}
\Bigg(\frac{\theta}{\sin \theta}-\lambda '\Bigg) \, ,
\ee 
Therefore it becomes clear that the undeterminacy manifests itself in the same fashion in both perturbative and non-perturbative treatments since they only differ by the effect of a shift in the integration momentum. Indeed as pointed out in \cite{JACPT} there is no apparent reason for these two approaches 
to produce different results. Our result is in consonance with those appearing in \cite{PER1} \cite{PER2} (see also \cite{CHAICHIANCPT},\cite{CHUNG2}) in the sense that the ambiguity stems from the undefined (regularization dependent) integral, $\int_k k_\mu k_\nu f(k^2) = c \, g_{\mu \nu} \int_k k^2 f(k^2)$. Indeed if we evaluated (\ref{undcpt}) using naively the symmetric limit ($c=1/4$) we would get
\bq
\Upsilon^0_{\mu \nu} &=& g_{\mu \nu}\int^{\Lambda}
\frac{d^4k}{(2 \pi)^4} \frac 1{(k^2-m^2)^2}
-g_{\mu \nu}\int^{\Lambda} \frac{d^4k}{(2 \pi)^4}
\frac {k^2}{(k^2-m^2)^3} \nonumber \\
&=& g_{\mu \nu}m^2\int \frac{d^4k}{(2 \pi)^4} \frac
1{(k^2-m^2)^3}=-\frac i{32 \pi^2}
g_{\mu \nu}.
\eq
which leads  to the result $3/(8 \pi^2)\epsilon^{\mu \nu \alpha \beta} p_\beta$ in the limit where $p^2=0$ \cite{JACPT}. In a position space regularization, this undeterminacy is expressed by its position space counterpart $\lim_{x \rightarrow 0} (x_\mu x_\nu)/x^2 = c g_{\mu \nu}$ \cite{CHAICHIANCPT}. Alternatively we could analyse such undeterminacy from the standpoint of the parametrizations as discussed in section \ref{s:int}. Since the arbitrariness is expressed by (\ref{undcpt}) with $\Upsilon_{\mu \nu}^0$ defined in (\ref{CR4L1}) we have
\bq
\Upsilon_{\mu \nu}^0 &=& g_{\mu \nu} ( I_{log}^{0\, \Lambda}(m^2) - 4 \, c \,  I_{log}^{1\, \Lambda}(m^2) ) \Rightarrow \nonumber \\
{\tilde{\Upsilon}}_{\mu \nu}^0 &=& g_{\mu \nu} \, b \Big(\ln \big( \frac{\Lambda_1^2}{m^2}\big) + \beta_1 - 4\, c\, \ln \big( \frac{\Lambda_2^2}{m^2}\big) - \beta_2\Big)
\eq
which is finite only if the symmetric limit ($c=1/4$) is taken. Therefore, if we choose such (general) parametrization to study the problem, the undeterminacies expressed by  $\lambda$, $\lambda '$ can be thought to come from the difference $\beta_1 - \beta_2$ which cannot be fixed by gauge invariance due the presence of the antisymmetric tensor.

\section{Adler-Bardeen-Bell-Jackiw Anomaly}
\label{s:abj}

The physical relevance of anomalies in quantum field theory was cleared up over $30$ years ago \cite{BELL},\cite{ADLER} and has been calculated in many regularization schemes \footnote{for an overview see \cite{BERTLMANN}}. Nevertheless it is very important to test IR in the triangle anomaly calculation. The reason is that such calculation constitutes a sort of ``acid test" for the consistency of a regularization framework (see \cite{BELLMEMO} for a nice account on this matter). Roughly speaking, from the perturbative standpoint, the anomaly manifests itself as an ambiguity represented by a local term due to underlying infinities of the diagrammatic calculation. Moreover such undeterminacy floats between the axial and vector channels, that is to say, the full (classical) tranversality for massless fermions is quantum mechanically broken. It is up to nature to decide how to make use of such freedom. On the other hand the most popular regularization prescriptions usually pick out tranversality on the vector currents to be fulfilled (e.g. Pauli-Villars, Zeta-function regularization) since vector gauge symmetry is fixed ab initio. A regularization framework which enables to display the anomaly evenly between the axial and vector Ward identities seems more appealing. 

In this section we study the Adler-Bardeen-Bell-Jackiw triangle anomaly and make explicit the role played by the CR and the momentum routing in the evaluation of the anomaly. 

To start we write the AVV triangle with arbitrary momentum routing, namely
\bq
T^{AVV}_{\mu \nu \alpha}&=&-\int_k  
\mbox{tr} \left\{ \gamma_{\mu}
(k\hspace{-2.2mm}/
+ k_1\hspace{-2.2mm}/
 -m )^{-1} \gamma_{\nu}
(k\hspace{-2.2mm}/
+ k_2\hspace{-2.2mm}/
 -m )^{-1} \gamma_{\alpha}
\gamma_{5}
(k\hspace{-2.2mm}/
+ k_3\hspace{-2.2mm}/
 -m )^{-1} \right\}\nonumber \\
&+& {\mbox{crossed\, diagram}}.
\eq
where the $k_i$'s are such that
\bq
k_2 - k_3 &=& p+q \nonumber \\
k_1 - k_3 &=& p \nonumber \\
k_2 - k_1 &=& q .
\label{emc}
\eq
We may parametrize the $k_i$'s to be consistent with (\ref{emc}) as
\bq
k_1&=&\alpha p+ (\beta -1) q ,\nonumber \\
k_2&=&\alpha p+\beta q, \nonumber \\
k_3&=&(\alpha -1)p + (\beta -1) q,
\label{park}
\eq
for general $\alpha$ and $\beta$. In the spirit of IR, we choose to write the (finite) differences between divergent integrals in terms of the CR (\ref{CR4Q1})-(\ref{CR4L2}) to yield
\bq
T_{\mu \nu \alpha}^{AVV}&=&\tilde {T}_{\mu \nu \alpha}^{AVV}+ \Big(i
\epsilon_{\mu \nu \beta \sigma}
\left[(k_2-k_1)^{\beta}+(k_3-k_1)^{\beta}\right]
g_{\alpha \rho} 
\Upsilon_0^{\rho \sigma}+ \nonumber \\
&-&i \epsilon_{\nu \alpha \beta \sigma}(k_2-k_3)^{\beta}
g_{\mu \rho} 
\Upsilon_0^{\rho \sigma} -i \epsilon_{\mu \alpha \beta
\sigma}(k_2-k_3)^{\beta} 
g_{\nu \rho} \Upsilon_0^{\rho \sigma} + \nonumber \\
&+&i \epsilon_{\mu \nu \alpha \sigma}
\left[(k_2+k_1)^{\beta}+(k_3+k_1)^{\beta}\right]
g_{\beta \rho} \Upsilon_0^{\rho \sigma} + {\mbox{crossed \, diagram}}\Big)
\label{tavv}
\eq
In the equation above $\tilde {T}_{\mu \nu
\alpha}^{AVV}$ is a finite, momentum routing independent quantity (it depends only on the differences expressed in (\ref{emc})), whose evaluation is sketched in the appendix. Notice however that we have terms which depend explicitly on the momentum routing. Using (\ref{park}) and $\Upsilon_{\mu \nu}^0 \equiv \lambda g_{\mu \nu}$ in (\ref{tavv}) we get
\bq
T_{\mu \nu \alpha}^{AVV}&=&\tilde {T}_{\mu \nu
\alpha}^{AVV} + 4i\lambda\left\{
\epsilon_{\mu \nu \alpha \beta}\left[ \alpha p^{\beta}
+ (\beta -1)q^{\beta}\right]-
\epsilon_{\mu \nu \alpha \beta}\left[ \alpha q^{\beta}
+ (\beta
-1)p^{\beta}\right]\right\}
\nonumber \\ &=& \tilde {T}_{\mu \nu \alpha}^{AVV} + 4i\lambda
(\alpha-\beta +1)
 \epsilon_{\mu \nu \alpha \beta}(p-q)^{\beta}.
\eq
Given that (see appendix)
\bq
p^\mu\tilde {T}_{\mu \nu \alpha}^{AVV}&=&-\frac
1{4\pi^2}\epsilon_{\mu \nu \alpha \beta}
p^\mu q^\beta \\
q^\nu\tilde{T}_{\mu \nu \alpha}^{AVV}&=&\frac
1{4\pi^2}\epsilon_{\mu \nu \alpha \beta}
p^\beta q^\nu \\
(p+q)^{\alpha}\tilde{T}_{\mu \nu \alpha}^{AVV}&=& 2 m T_{\mu
\nu},
\eq
where $T_{\mu \nu}$ is the usual vector-vector-pseudoscalar triangle amplitude we can write the Ward identities as
\bq
p^{\mu}T_{\mu \nu \alpha}^{AVV}&=& \left\{-\frac 1{4\pi^2} -
4i\lambda 
(\alpha -\beta +1)\right\} \epsilon_{\mu \nu \alpha \beta}p^{\mu}q^{\beta} \nonumber \\
q^{\nu}T_{\mu \nu \alpha}^{AVV}&=&\left\{\frac 1{4\pi^2}+
4i\lambda 
(\alpha -\beta +1 )\right\}
\epsilon_{\mu \nu \alpha \beta}q^{\nu}p^{\beta} \nonumber \\
(p+q)^{\alpha}T_{\mu \nu \alpha}^{AVV}&=&2mT_{\mu
\nu}-8i\lambda (\alpha -\beta +1)
\epsilon_{\mu \nu \alpha \beta}p^{\alpha}q^{\beta}.
\label{wi}
\eq
Now $\lambda$ is a local arbitrary parameter and the routing labeled by $\alpha$,$\beta$ can assume any value. In particular we could redefine $\lambda (\alpha -\beta +1) \rightarrow \lambda '$. We finally write the Ward identities as
\bq
p^{\mu}T_{\mu \nu \alpha}^{AVV}&=&-\frac 1{8\pi^2}(a + 1) 
\epsilon_{\mu \nu \alpha
\rho}p^{\mu}q^{\rho} \\
(p+q)^{\alpha}T_{\mu \nu \alpha}&=&2mT_{\mu \nu}-\frac
1{4\pi^2}(a-1) 
\epsilon_{\mu \nu \alpha
\rho}p^{\alpha}q^{\rho}.
\eq
where we defined $4 i \lambda ' \equiv (a - 1)/(8\pi^2)$. Hence we clearly see that our approach correctly displays the anomaly: the choice of $a$, which is ultimately related to the arbitrary value of the CR and the momentum routing, enables 
us to fix tranversality either in the vector or the axial Ward identity. The reason is that the choice of the arbitrary parameters were left till the very end of the calculation to be fixed. This corresponds to fixing  the ratio of renormalization scales in DFR.


\section{Conclusions and Outlook}

We have successfully tested an implicit regularization framework (IR) that works directly in momentum space to study CPT violation in an extended version of $QED_4$ and the triangle chiral anomaly, where regularization plays a delicate role. The main purpose of our program is to construct a consistent regularization approach to dimension specific models, such as chiral, topological and supersymmetric models since we work in the (integer) space-time dimension where the theory is defined and no explicit change in the Lagrangian of the theory is effected. Feynman diagram calculation in dimension-specific models are often plagued by spurious anomalies especially beyond the one-loop order. 

Yet a constrained version of IR is more practical from the calculational standpoint and more convenient as it appears to fix gauge invariance from the start, we have seen that one should be careful in studying problems in which (an odd number of) parity violating objects appear. As we have seen in the CPT violation problem discussed in section \ref{s:cpt} gauge invariance does not fix the undeterminacy. Another instance where the CR should be left arbitrary is the Chiral Schwinger Model discussed in \cite{PRD}: should we choose to work with the constrained version we would have obtained a wrong mass spectrum, which is known from non-perturbative calculations to be essentially undetermined in a range of values dictated by unitarity. Had we chosen to set the CR equal to zero for the triangle anomaly calculation discussed in section \ref{s:abj} we would inforce momentum routing invariance. 
Despite loosing the democracy between the AWI and VWI in what concerns the symmetry breaking
by setting $\lambda=0$ in (\ref{wi}) and consequently violating the VWI we can still recourse to finite renormalization in a similar fashion as discussed in \cite{GERS}. By redefining a physical amplitude as $T_{\mu \nu \alpha}^{phys} = T_{\mu \nu \alpha} - T_{\mu \nu \alpha} (0)$, where $T_{\mu \nu \alpha}(0) = 1/(4 \pi^2) \epsilon_{\alpha \mu \nu \lambda} (p - q)_\lambda$ it is easy to see that we restore the VWI whereas the anomaly goes to the AWI. For the counterpart in DFR please see \cite{PV3}.                       

IR may be also applied to Chern-Simons-Matter (CSM) model calculations. The latter involves a $3$-dimensional Levi-Civitta tensor which is just the analogue of the $\gamma_5$ matrix in $3$-dimensions. Usually CSM are evaluated using a combination of High-Covariant Derivatives which are introduced in the Lagrangian and a modified version of DR (t'Hooft Veltman rules). This turns the propagators very complicated and unpractical for computation beyond the one loop order \cite{KUNST}. Finally it has been recently shown \cite{GC} that IR can be used to construct an algebraic proof of renormalizability for $\varphi^3_6$-theory in an alternative fashion to BPHZ method. The next step would be to implement this approach for a gauge field theory as well as work with IR in more symmetric models and at higher loop orders \cite{WIP}.

\section*{Appendix: $\tilde T^{AVV}_{\lambda \mu \nu}$ calculation} 

As $\tilde T^{AVV}_{\lambda \mu \nu}$  is momentum routing independent, define
$T^{AVV}_{\lambda \mu \nu}$ for the routing  $k_1=0, k_2=q, k_3=-p$:

\begin{eqnarray}
T^{AVV}_{\lambda \mu \nu} & = & i  \int_k
{\mbox{tr}} \{\gamma^{\lambda} \gamma_5 (\not k - \not p - m)^{-1}
\gamma^{\mu} (\not k-m)^{-1} \gamma^{\nu} (\not k + \not q - m)^{-1}\}\nonumber\\
& = & \tilde T^{AVV}_{\lambda \mu \nu} + {\mbox{ terms\, multiplying}}\,\, \Upsilon 's.\nonumber\\
\tilde T^{AVV}_{\lambda \mu \nu}
&\equiv &
\{\epsilon_{\lambda \mu \nu \omega} (p_{\omega} - q_{\omega}) F_1
(p,q)
\nonumber\\ & + & p_{\omega} q_{\phi} \{[\epsilon_{\lambda \nu
\omega \phi}
q^{\mu} + \epsilon_{\lambda \mu \omega \phi} q^{\nu}]F_2 (p,q)
\nonumber\\
& + & [\epsilon_{\lambda\nu \omega \phi} p^{\mu} +
\epsilon_{\lambda \mu \omega \phi} p^{\nu}] F_3 (p,q)\nonumber\\
& + & \epsilon_{\mu \nu \omega \phi} [ p^{\lambda} F_4 (p,q) +
q^{\lambda}
F_5 (p,q)]\}\nonumber\\ & + & \epsilon_{\lambda \mu \nu \omega}
[ p_{\omega}
F_6 (p,q) - q_{\omega} F_7 (p,q)]\} .
\end{eqnarray}
with
\begin{equation}
F_1 (p,q) = -{1\over (4\pi^2)} \left[{Z_0\over 4} ((p+q)^2; m^2) -
{1\over 4} - {m^2 \xi_{00}(p,q)\over 2} + {q^2 \xi_{01}(p,q) +
p^2 \xi_{10}(p,q)\over
4}\right]
\end{equation}
\begin{equation}
F_2 (p,q) = {1\over (4\pi^2)} [\xi_{01}(p,q) - \xi_{02}(p,q) -
\xi_{11}(p,q)]
\end{equation}
\begin{equation}
F_3 (p,q) =  {1\over (4\pi^2)} [\xi_{11}(p,q) + \xi_{20}(p,q) -
\xi_{10}(p,q)]
\end{equation}
\begin{equation}
F_4 (p,q) = - {1\over (4 \pi^2)} [\xi_{11}(p,q) + \xi_{10}(p,q) -
\xi_{20}(p,q)]
\end{equation}
\begin{equation}
F_5 (p,q) = - {1\over (4\pi^2)} [\xi_{11}(p,q) + \xi_{01}(p,q) -
\xi_{02}(p,q)]
\end{equation}
\begin{equation}
F_6 (p,q) = - {1\over (4\pi^2)}\left[-{Z_0(p^2; m^2)\over 4} -
{(p+q)^2
\over 2} \xi_{10}(p,q) + 4m^2\xi_{00}(p,q) \right]
\end{equation}
\begin{equation}
F_7 (p,q) = - {1\over 4\pi^2} \left[-{Z_0 (q^2; m^2)\over 4} -
{(p+q)^2
\over 2} \xi_{01}(p,q) + 4m^2\xi_{00}(p,q)\right]
\end{equation}
where the functions $\xi_{nm}(p,q)$ are defined as \cite{PHD}

\begin{equation}
\xi_{n m}(p,q) = \int^1_0 \, dz \int^{1-z}_0
\, dy {z^n y^m\over Q (y, z)}
\end{equation}
with
\begin{equation}
Q = (y, z) = p^2 y (1-y) + q^2 z (1-z) - m^2 + 2 \, p.q\,  yz.
\end{equation}
At the origin $\xi_{n m}(0,0)=-\frac{1}{2m^2}$. The functions $Z_k$ are defined as
\begin{equation}
Z_k (p^2; m^2) = \int^1_0 dz \, z^k \ln \, \left({p^2 z (1-z)-m^2
\over - m^2}\right)
\label{zk}
\end{equation}
The following relations between the functions $Z_k$ and $\xi_{mn}$ can be easily checked and greatly simplifies the calculation of the $T^{AVV}_{\lambda \mu \nu}$ and the Ward identities. 
\begin{equation}
q^2 \xi_{01}(p,q) - p.q \xi_{10}(p,q) = {1\over 2} \{Z_0 (q^2;
m^2) -
Z_0 (p.q; m^2) + p^2 \xi_{00}(p,q)\}
\end{equation}
\begin{equation}
q^2\xi_{11}(p,q) - p.q\xi_{20}(p,q) = {1\over 2} \left\{{-Z_0(p +
q)^2;m^2)
\over 2}
+ {Z_0(p^2; m^2)\over 2} + q^2 \xi_{10}(p,q)\right\}
\end{equation}
\begin{equation}
q^2\xi_{02}(p,q) - p.q\xi_{11}(p,q) = {1\over
2}\left\{-\left[{1\over 2} + m^2
\xi_{00}(p,q)\right] + {p^2\over 2} \xi_{10} (p,q) + {3q^2\over 2}
\xi_{01}(p,q)\right\}
\end{equation}
\begin{equation}
p^2\xi_{20}(p,q) - p.q\xi_{11}(p,q) = {1\over 2}
\left\{-\left[{1\over 2}
+ m^2 \xi_{00}(p,q)\right] + {q^2\over 2} \xi_{01}(p,q) +
{3p^2\over 2} \xi_{10}(p,q)\right\}
\end{equation}
\begin{equation}
p^2\xi_{11}(p,q) - p.q\xi_{02}(p,q) = {1\over 2} \left\{-{1\over 2}
Z_0((p+q)^2;m^2) + {1\over 2} Z_0(q^2;m^2) + p^2
\xi_{01}(p,q)\right\} \, .
\end{equation}
Adding up the crossed diagram we can readily see that
\bq
p^\mu\tilde {T}_{\mu \nu \alpha}^{AVV}&=&-\frac
1{4\pi^2}\epsilon_{\mu \nu \alpha \beta}
p^\mu q^\beta \\
q^\nu\tilde{T}_{\mu \nu \alpha}^{AVV}&=&\frac
1{4\pi^2}\epsilon_{\mu \nu \alpha \beta}
p^\beta q^\nu \\
(p+q)^{\alpha}\tilde{T}_{\mu \nu \alpha}^{AVV}&=& 2 m T_{\mu
\nu}. \quad\quad\quad\quad\quad {\mbox{qed}}
\eq

\section*{Acknowledgements}
The authors are grateful to Dr. W. F. Chen and Dr. Manolo P\'erez-Victoria for enlightening discussions. This work is supported by FCT/MCT/Portugal under the grant numbers PRAXIS XXI-BPD/22016/99, PRAXIS XXI/BCC/4301/94, PRAXIS/C/FIS/12247/98, POCTI/1999/FIS/35309, CNPq/Brazil and Fapemig/Brazil.


\end{document}